# Hopkinson peak and superparamagnetic effects in $BaFe_{12-x}Ga_xO_{19}$ nanoparticles

S. H. Mahmood

Physics Department, The University of Jordan, Amman, Jordan,
E-mail: s.mahmood@ju.edu.jo,
Phone: (962) 796709673.

I. Bsoul

Physics Department, Al al-Bayt University, Mafraq 130040, Jordan,

E-mail: Ibrahimbsoul@yahoo.com , Phone: (962 2) 6297000 (ext. 3588), Fax: (962 2) 6297031.

**Abstract**

In this article, the thermomagnetic properties of a system of Ga-substituted barium hexaferrite nanoparticles ($BaFe_{12-x}Ga_xO_{19}$) prepared by ball milling were investigated. The thermomagnetic curves for the samples with *x* ranging from 0.0 to 1.0 exhibited sharp peaks with high magnetization just below $T_C$ (Hopkinson peaks). The height of the peak for our samples was similar or larger than previously observed or calculated values. Theoretical treatment of the experimental data demonstrated that the peaks are due to the effect of superparamagnetic relaxations of the magnetic particle. This effect was confirmed by hysteresis measurements at, and just below the temperature at which the peak occurred. Consequently, the particle diameters were calculated from the experimental data using a theoretical model based on the superparamagnetic behavior of a system of uniaxial, randomly oriented, single domain, non-interacting particles. The calculated diameters of 11 - 26 nm are less than the physical diameters determined from TEM measurements. The factors responsible for the low calculated values are discussed.

# 1. Introduction

Barium hexaferrite $BaFe_{12}O_{19}$ (BaM) possess interesting properties such as large saturation magnetization, high coercivity, high Curie temperature, large uniaxial magnetic anisotropy and chemical stability. These materials have been investigated due to their importance for both fundamental research and technological applications in permanent magnets, high-density magnetic recording, magneto-optics and microwave devices [1-6]. Different techniques have been used to prepare and characterize hexaferrite particles [7-13]. The magnetic properties of these materials have been tuned by substitution of Fe by different magnetic and nonmagnetic cations, and intensively investigated by different techniques [14-20].

The magnetization of certain hexaferrites exhibits a peak (Hopkinson peak) near $T_C$ in the thermomagnetic curve in a weak applied magnetic field. Popov and Mikhov explained this effect using Stoner-Wohlfarth model for magnetically stable, single-domain (SD), randomly oriented particles [21]. According to this model, the magnetization is given by:

$$M_{SD} = \frac{2}{3} p \frac{M_s(T)}{H_a(T)} H \qquad (1)$$

where $p$ is the packing fraction of the powder, $M_s(T)$ is the bulk saturation magnetization, and $H_a(T)$ is the anisotropy field at temperature $T$. The explanation was based on the argument that the competition between the increase in magnetization due to the decrease in the anisotropy field, and the decrease in magnetization due to the decrease in saturation magnetization as the temperature increases may result in a peak in the magnetization. Later, the effect of the demagnetizing field was taken into consideration in explaining the origin of Hopkinson peak [22]. Using a completely different approach, the origin of the peak was explained by the superparamagnetic behavior of magnetic particles which do not exhibit a peak as a result of the variations of the saturation magnetization and anisotropy field of the material [23].

In the present work we prepared a system of BaM nanoparticles doped with Gallium, and investigated its structural and magnetic properties. The Hopkinson peak height was analyzed in terms of the superparamagnetic relaxation processes of the particles and compared with the experimental data to arrive at a conclusion concerning the mechanism responsible for the peak, and the particle size distribution of the synthesized powders.

## 2. Experimental procedures

$BaFe_{12-x}Ga_xO_{19}$ powders with $x$ ranging from 0.0 to 1.0 were prepared using high energy ball milling and appropriate heat treatment. The samples were characterized using XRD and transmission electron microscopy (TEM), and the magnetic measurements were performed using a vibrating sample magnetometer (VSM). For further details on the experimental procedures the reader is referred to our earlier publication [24].

## 3. Results and discussion

Fig. 1 shows the XRD patterns of samples of $BaFe_{12-x}Ga_xO_{19}$ along with the standard pattern (JCPDS: 043-0002) for hexagonal barium ferrite ($BaFe_{12}O_{19}$) with space group P6$_3$/*mmc*. No secondary phases were detected in the diffraction patterns indicating the formation of a pure phase with variations in the lattice parameters less than 0.1%.

The average crystallite size was determined using Scherrer formula [25],

$$D = \frac{k\lambda}{\beta\cos\theta}, \qquad (2)$$

where $D$ is the crystallite size, $k$ the Scherrer constant (= 0.94), $\lambda$ the wavelength of radiation (1.54056 Å), $\beta$ the peak width at half maximum measured in radians, and $\theta$ the peak position. The average crystallite size for the pure and doped samples ranges from 37 nm to 45 nm.

TEM images of representative samples are shown in Fig. 2. The average particle size for the pure sample is (42 ± 13) nm, and for the sample with $x = 1.0$ is (41 ± 13) nm. These values indicate that the synthesized powders consist of single domain magnetic nanoparticles with a relatively narrow particle size distribution.

The initial magnetization of the pure sample was checked and found to increase slowly at low fields followed by a rapid increase at higher fields [24]. This behavior is typical for randomly oriented single domain magnetic particles. The variations of the saturation magnetization and coercivity with $x$ are shown in Fig. 3, and their values together with Curie temperatures for the samples are listed in Table 1. The behavior of the saturation magnetization and coercivity indicates that Ga ions replace Fe ions at both spin-up and spin-down sites. The small initial drop in coercivity and slow decrease in saturation magnetization for $x$ values up to 0.2 are consistent with the substitution of Ga at spin-up $2a$ and spin-down $4f_1$ sites with preference for occupying $2a$ sites. This is consistent with the substitution of small amounts of Ti-Ru at these sites as confirmed by Mossbauer spectroscopy [26]. For higher $x$ values, spin-up $12k$ sites which contribute negatively to the anisotropy field start getting occupied by Ga ions, leading to the observed increase in coercivity and decrease in saturation magnetization. The change in behavior of $M_s$ at $x = 0.6$ suggests that beyond this value, the fraction of Ga ions substituting Fe ions at spin-down sites remains constant at a value of 0.2, where the remaining fraction substitute Fe ions at spin-up $12k$ sites. This substitution would lead to the observed 5% drop in $M_s$ at $x = 0.6$, and the 15% drop at $x = 1.0$. The remanence ratio $M_{rs} = M_r/M_s$ is ~ 0.5 (Table 1), which is consistent with the theoretical value for a system of uniaxial, single domain, randomly oriented particles.

Fig.4 shows the thermomagnetic curves as a function of temperature for the samples at a constant applied field of 100 Oe. All curves exhibit sharp pronounced Hopkinson peaks just below $T_C$, and the height of the peak relative to the minimum magnetization (RPH) for all samples is shown in Table 2. The sharpness of the peaks indicates a narrow particle size distribution. The relative peak heights for our samples are similar or higher than the observed value of about 10 for Co-Ti substituted sample, and the calculated value of about 8 based on superparamagnetic relaxation [23]. To investigate the origin of the peak, a sample is prepared from bulk Barium hexaferrite (Aldrich made) powder of grain size ~ 0.5 - 2 μm (with small fraction of smaller particles), and $M_s$ for the sample are measured against temperature. The anisotropy field $H_a$ for the sample is determined from the switching field distribution evaluated by differentiating the reduced DC demagnetization curve [24]. The magnetization is then calculated from these values by adopting Stoner-Wohlfarth model for SD blocked particles (eq. (1)) and the results are shown in Fig. 5. The figure shows only a very small rise in the magnetization below $T_C$ which is insignificant compared with the peak heights observed for our samples. Further, the magnetization is measured versus $T$ for the bulk sample (Fig. 5). The measured magnetization shows a relatively small sharp peak with relative height of 1.8, and a behavior similar to that of the calculated magnetization in the temperature range below the peak. The higher values of the measured magnetization in this temperature range could be associated with the effects of interparticle interactions, or the presence of multidomain particles in the sample, which are not accounted for in Stoner-Wohlfarth theory. Thus the observed sharp peaks in the thermomagnetic measurements on our samples cannot be due to the temperature dependences of the saturation magnetization and anisotropy field. The small peak observed for the bulk sample could be associated with a small fraction of superparamagnetic particles in the bulk powder, or with the temperature dependence of the saturation magnetization and the anisotropy field [21, 22]. However, the relative height of this peak cannot account for the large observed peaks in our samples. Accordingly, we are led to believe that the observed Hopkinson peaks in our synthesized samples are associated with the superparamagnetic relaxations of the particles in the samples.

The magnetic relaxations of the particles are further confirmed by measuring the hysteresis loops at the peak temperature and at lower temperatures. Fig. 6 shows the loops for the sample with $x = 0$, which show superparamagnetic behavior with almost zero coercivity at the peak temperature, and the appearance of coercivity at the temperature of the minimum magnetization just before the rise of the peak (at 460 ºC). The increase in coercivity as the temperature is lowered is a consequence of the gradual blocking of the particles as the temperature is lowered. All samples show similar behavior, as illustrated in Fig. 7 for the sample with $x = 0.4$.

Assuming that the volumes of the superparamagnetic particles have a flat top distribution between $V_1$ and $V_2$, the upper and lower limits of the particle volumes can be calculated following the model calculation in [27]. Following this model, the upper limit is calculated from the initial susceptibility given by:

$$\chi = \frac{dM}{dH} = \frac{M_s^2 V_2}{3kT} \qquad (3)$$

where $M_s$ is the saturation magnetization at temperature $T$. The lower limit is calculated from the slope of the magnetization versus $1/H$ in the high field region, which is given by [27]:

$$M = M_s - \frac{kT}{V_1 H} \qquad (4)$$

The initial susceptibility, saturation magnetization, and the slope of *M* vs. 1/*H* are determined from the magnetization curve (Fig. 8) for each sample at the peak temperature where the sample behaves superparamagnetically.

The calculated particle diameters, assuming spherical particles, are listed in Table 2. The particle diameters range from 11 nm – 26 nm, which are lower than previously reported results [24]. Differences between particle diameters evaluated from the magnetic data and the previously reported values could be due to deviation of the particle size distribution from the assumed flat top distribution, and to interparticle interactions. However, these effects are possibly not enough to account for the observed reduction of more than 50%. A number of reasons, in addition to the assumptions of the theory, could be responsible for the low calculated values. Firstly, the calculated volume is the volume of the magnetic core of the particle, and a nonmagnetic shell (dead layer) could be surrounding the particles, which would give smaller particle sizes than the physical sizes. Secondly, the particles could be platelets in shape rather than spherical as suggested by the TEM images. A rough estimate of the platelet thickness could be obtained by assuming that the observed physical diameter is that of a cylindrical platelet of volume equals to that calculated from the magnetic data. In this case, the thickness of the platelets is found to be between 1.5 nm and 4.0 nm. The lower limit of the thickness is smaller than the lattice parameter c, which is evidence that the magnetic volume is an under estimate of the physical volume.

## 4. Conclusions

Barium hexaferrite nanoparticle systems doped with different concentrations of Ga prepared by ball milling exhibit single hexagonal phase with crystallite size ranging between 37 and 45 nm. The magnetic measurements as a function of temperature exhibit sharp peaks with high relative magnetization which cannot be explained on the basis of Stoner-Wohlfarth model for an assembly of randomly oriented, non-interacting, single-domain particles. The magnetization curves at the peak temperatures for all samples are consistent with the behavior of a system of superparamagnetic particles, which are blocked at slightly lower temperature, indicating narrow superparamagnetic particle size distribution. The calculated particle diameters for these samples are between 11 nm and 26 nm. These values suggest that the samples consist of single magnetic domain particles.

**Acknowledgement**

This work is supported by a generous grant from the Scientific Research Support Fund in Jordan under grant number (S/21/2009). The work is accomplished during a sabbatical leave provided by Yarmouk University to one of the authors (S.H.M.) which was spent at Al al-Bayt University. The Authors would like to thank Munir Khdour, Yarmouk University, for his technical assistance in electron microscopy.

**Figure Captions**

Fig. 1 Standard JCPDS pattern for M-type hexagonal barium ferrite (file no: 043-0002) and XRD patterns of $BaFe_{12-x}Ga_xO_{19}$ with different doping concentration.

Fig. 2 TEM images of $BaFe_{12-x}Ga_xO_{19}$, a) $x = 0.0$, b) $x = 1.0$.

Fig. 3 Saturation magnetization and coercivity variations with $x$ for $BaFe_{12-x}Ga_xO_{19}$

Fig. 4 Thermomagnetic curves of $BaFe_{12-x}Ga_xO_{19}$.

Fig. 5 Experimental and calculated thermomagnetic curves of bulk barium hexaferrite.

Fig. 6 Hysteresis loops for the sample with $x = 0$ at different temperatures.

Fig. 7 Hysteresis loops for the sample with $x = 0.4$ at different temperatures.

Fig. 8 Magnetization curve for the sample with $x = 0.4$ at peak temperature (414 °C).

**Table Captions**

Table 1: Coercivity, Saturation magnetization, remanence ratio ($M_{rs} = M_r/M_s$), and Curie temperature for Ga-substituted hexaferrite samples

Table 2: Hopkinson peak temperature ($T_p$), saturation magnetization at the peak temperature, particle diameters, and relative peak height (*RPH*) for Ga-substituted hexaferrite samples.

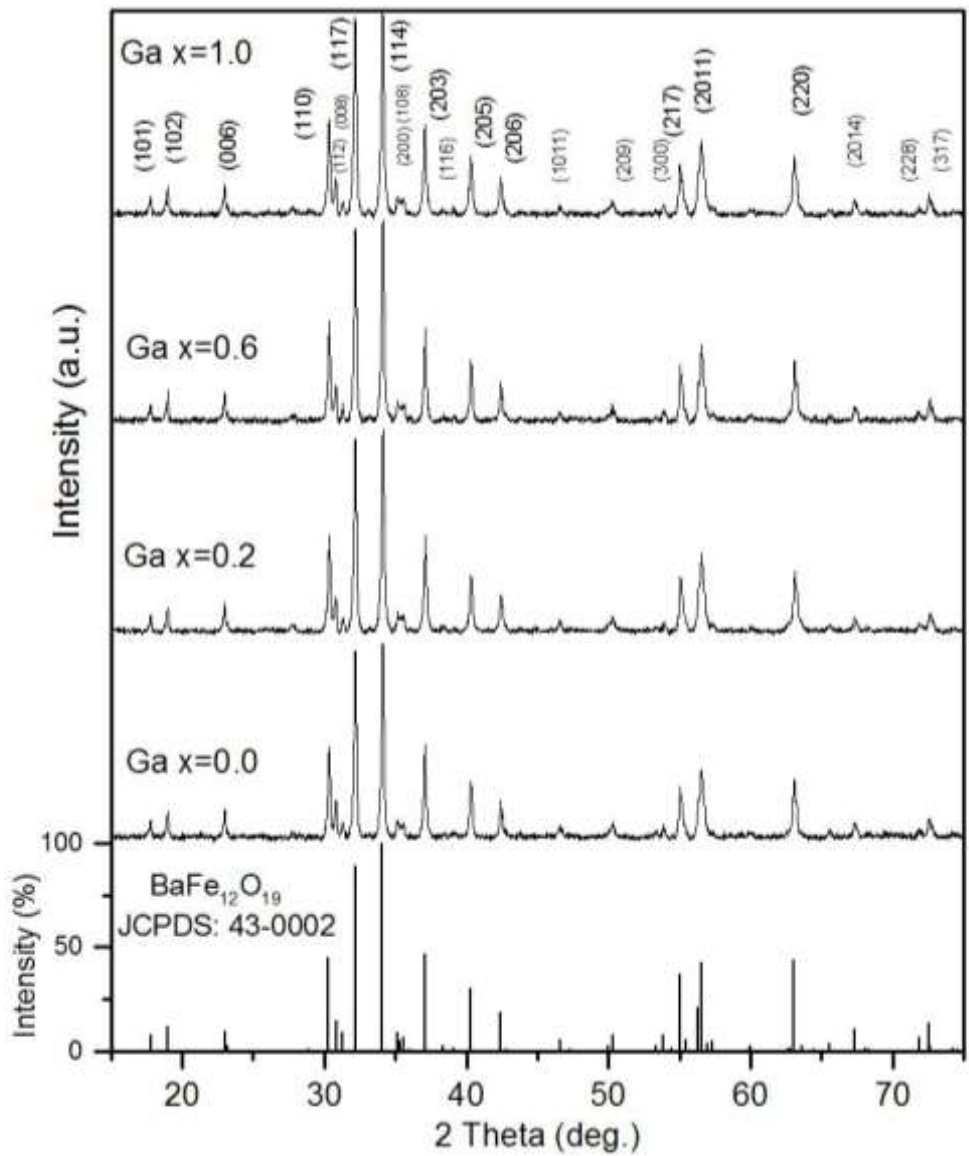

Fig. 1/ Mahmood and Bsoul

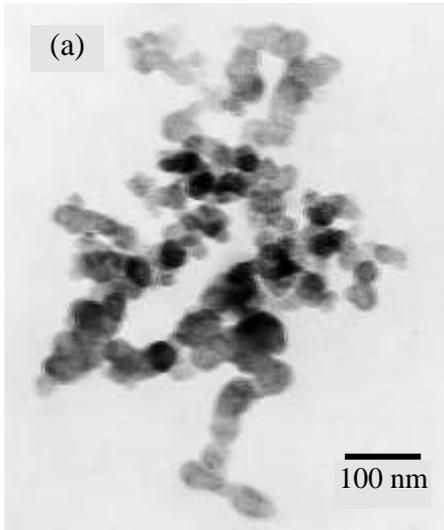 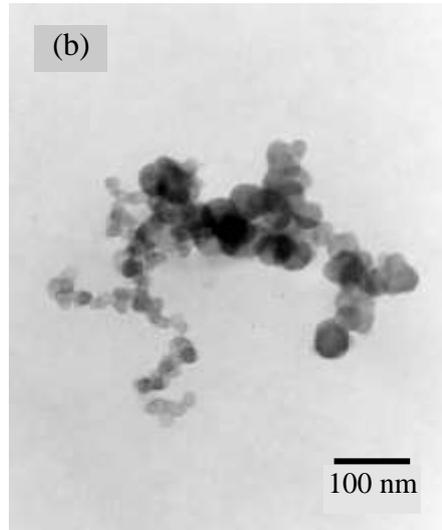

Fig. 2/ Mahmood and Bsoul

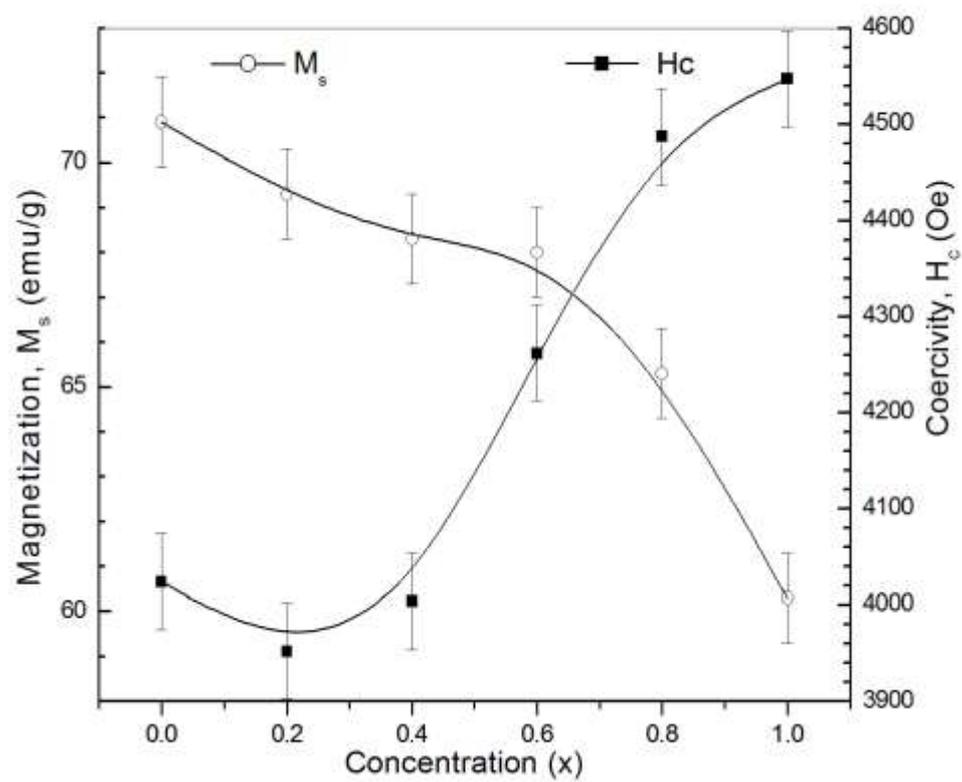

Fig.3/ Mahmood and Bsoul

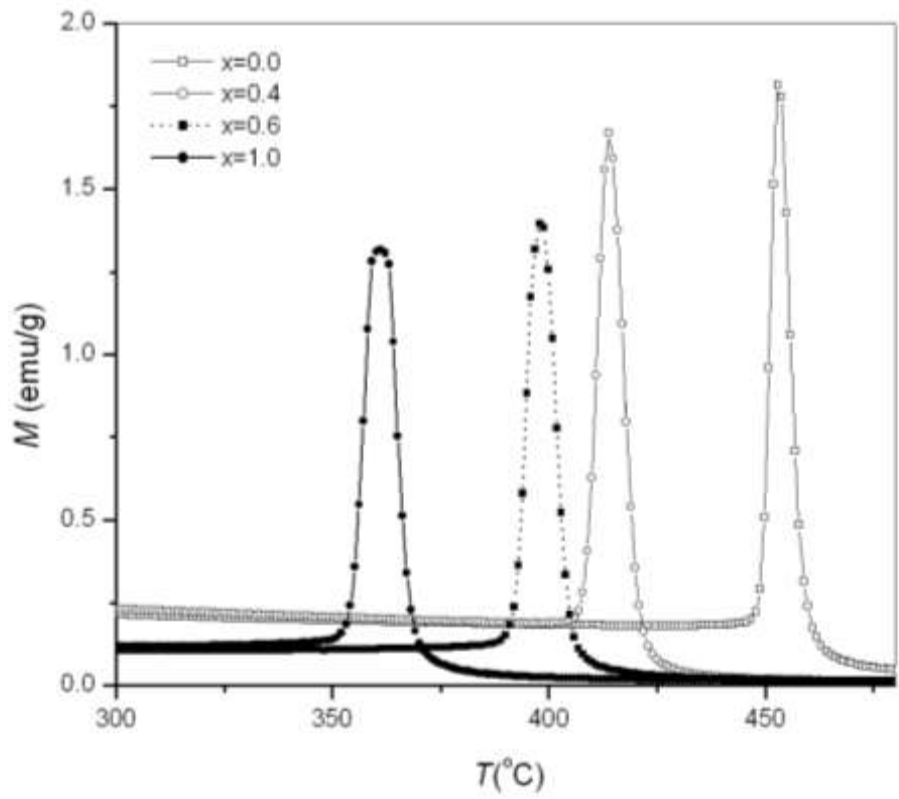

Fig.4/ Mahmood and Bsoul

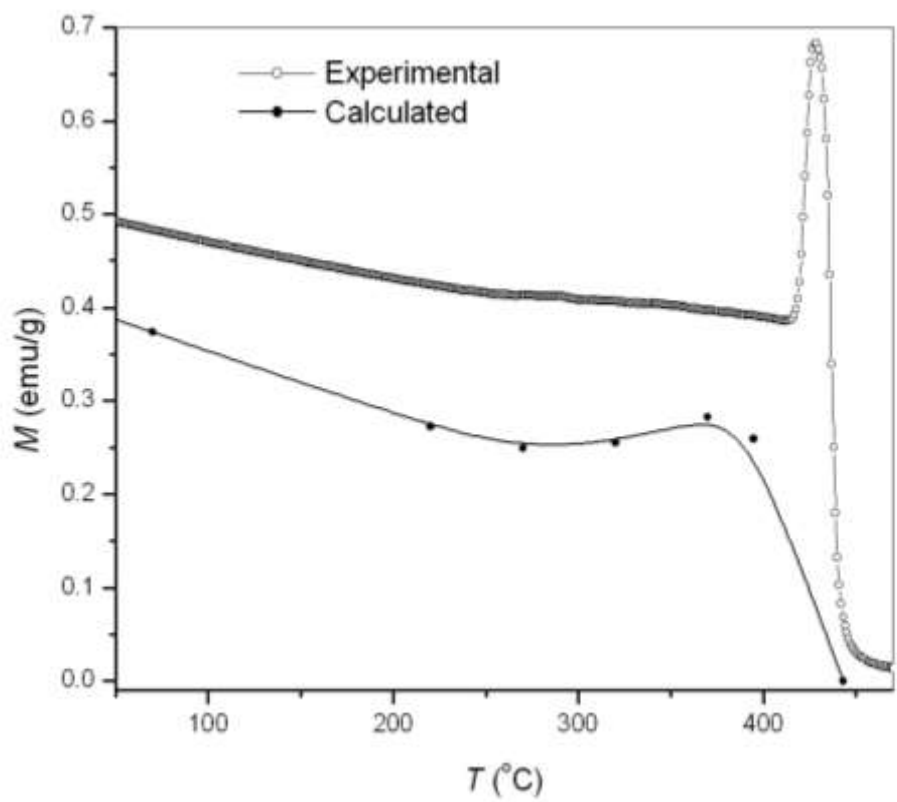

Fig. 5/ Mahmood and Bsoul

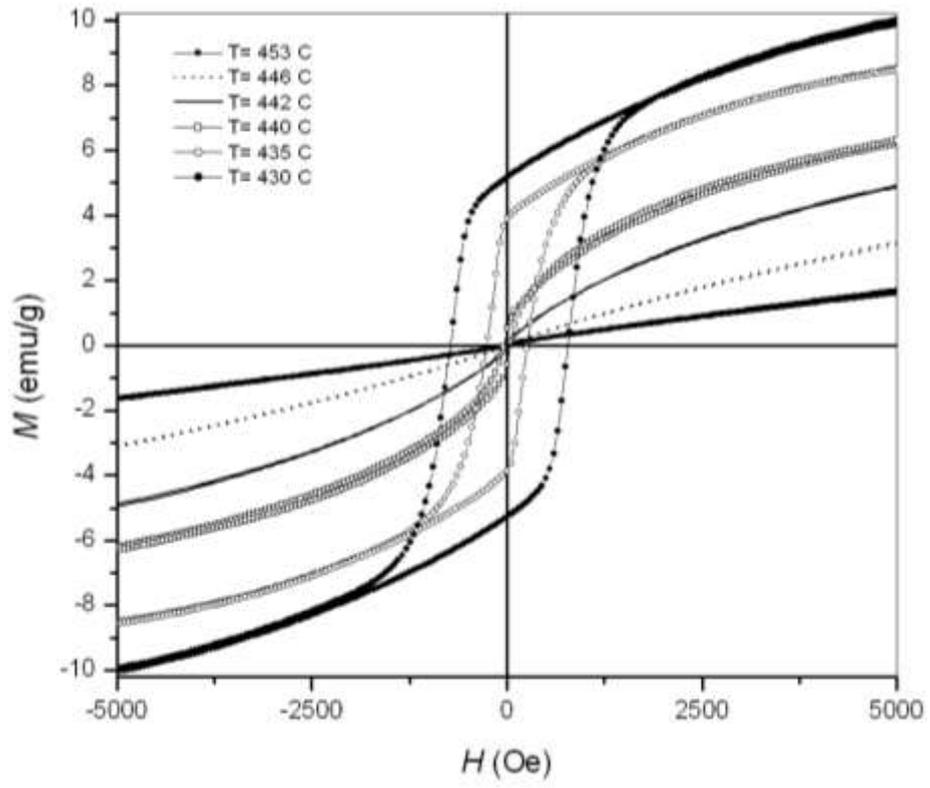

Fig. 6/ Mahmood and Bsoul

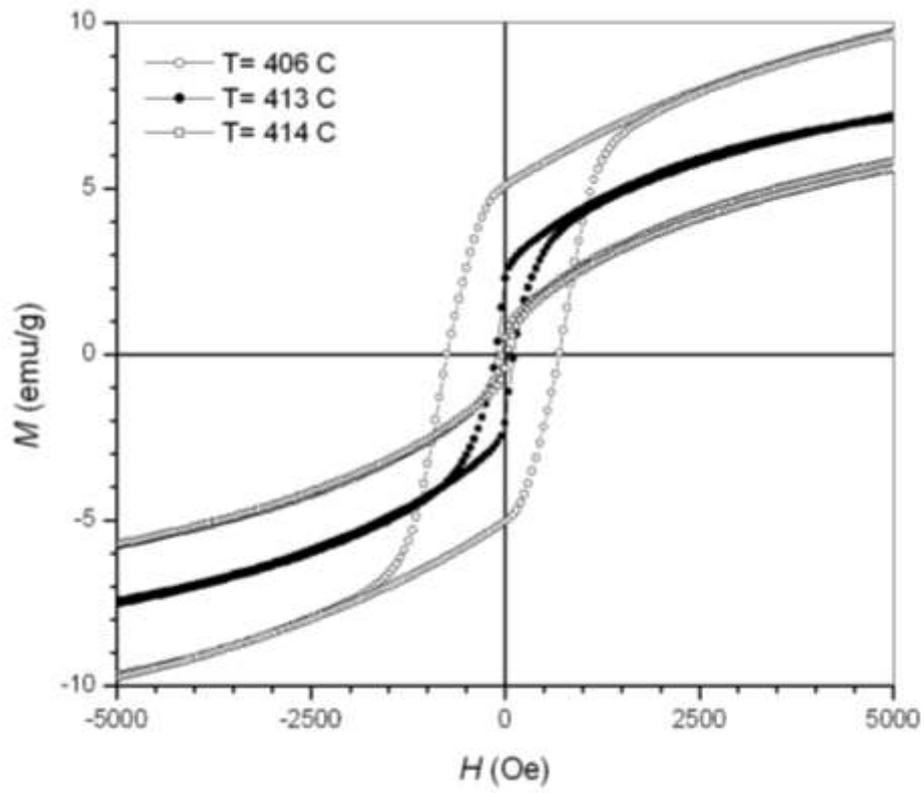

Fig. 7/ Mahmood and Bsoul

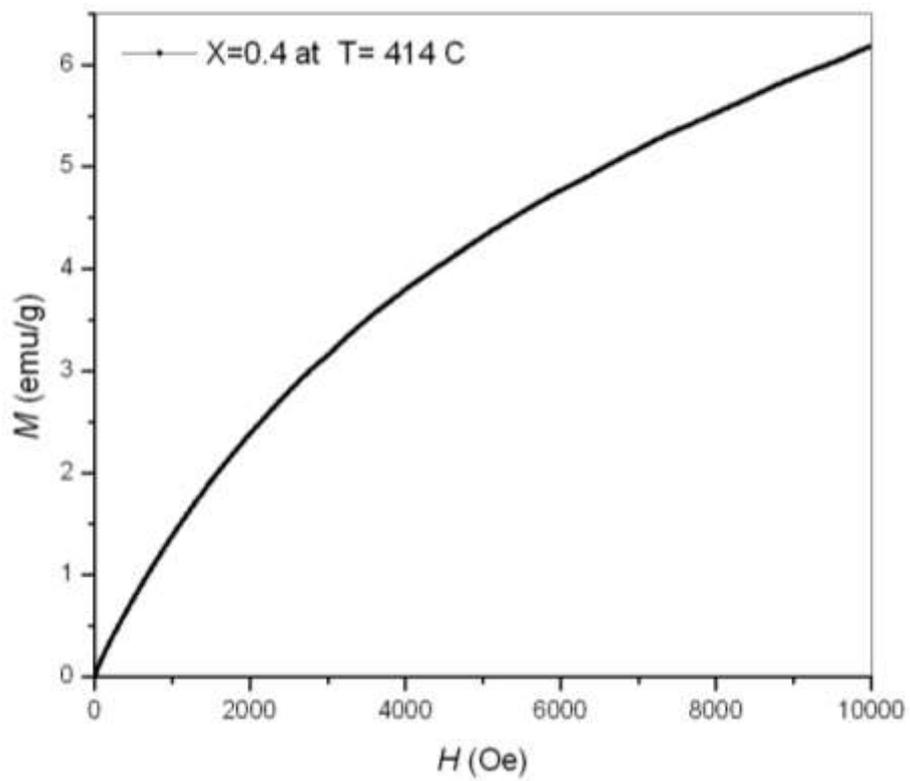

Fig. 8/ Mahmood and Bsoul

Table 1/ Mahmood and Bsoul

| $x$ | $H_c$ (kOe) | $M_s$ (emu/g) | $M_{rs}$ | $T_c$ (°C) |
|---|---|---|---|---|
| 0.0 | 4.02 | 70.9 | 0.523 | 460 |
| 0.2 | 3.95 | 69.3 | 0.519 | 430 |
| 0.4 | 4.00 | 68.3 | 0.518 | 420 |
| 0.6 | 4.26 | 68.0 | 0.519 | 400 |
| 0.8 | 4.49 | 65.3 | 0.517 | 385 |
| 1.0 | 4.55 | 60.3 | 0.514 | 370 |

Table 2/ Mahmood and Bsoul

| $x$ | $T_p$ (°C) | $M_s(T_p)$(emu/g) | $D_1$(nm) | $D_2$(nm) | RPH |
|---|---|---|---|---|---|
| 0.0 | 453 | 5.83 | 12 | 14 | 10.8 |
| 0.2 | 428 | 7.12 | 12 | 13 | 9.3 |
| 0.4 | 414 | 10.07 | 12 | 22 | 8.9 |
| 0.6 | 398 | 9.72 | 13 | 26 | 12 |
| 0.8 | 382 | 8.12 | 12 | 20 | 10.5 |
| 1.0 | 362 | 8.72 | 11 | 14 | 10.8 |